\documentstyle[twocolumn,prl,aps]{revtex}
\title{Renormalization group approach to nonextensive statistical mechanics}
\author{Renio S. Mendes$^{1,2}$ and Constantino Tsallis$^{1,3}$}
\address{$^1$Department of Physics, University of North Texas, P.O. Box 311427, Denton, Texas 76203-1427, USA \\
$^2$Departamento de Fisica, Universidade Estadual de Maringa, Avenida Colombo 7590, 87020-900 Maringa-PR, Brazil\\
$^3$Centro Brasileiro de Pesquisas Fisicas,
 Rua Xavier Sigaud 150, 22290-180 Rio de Janeiro-RJ, Brazil\\
rsmendes@dfi.uem.br, tsallis@unt.edu, tsallis@cbpf}

\draft

\begin{document}

\maketitle

\begin{abstract}
We analyze a simple classical Hamiltonian system within the hypothesis
of {\it renormalizability} and {\it isotropy} that essentially led Maxwell
 to his ubiquitous Gaussian distribution of velocities. We show that
  the equilibrium-like {\it power-law} energy distribution emerging
  within nonextensive statistical mechanics satisfies these hypothesis,
   {\it in spite of not being factorizable}. A physically satisfactory
   renormalization group emerges in the $(q, T_q)$ space, where $q$
   and $T_q$ respectively are the entropic index characterizing
    nonextensivity, and an appropriate temperature. This scenario
     enables the conjectural formulation of the one to be expected for
     $d$-dimensional systems involving long-range interactions (e.g., a
      classical two-body potential
       $\propto r^{-\alpha}$ with $0 \le \alpha/d \le 1$). As a corollary,
        we recover a quite general expression for the classical principle
        of equipartition of energy for arbitrary $q$.
\end{abstract}



If we allow ourselves to use a contemporary vocabulary, we may say that
the essential hypothesis that led James Clerk Maxwell to the classical
distribution of velocities that is named after him were two, namely
{\it renormalizability} and {\it isotropy}. Let us be more specific.
If we have the one- and two-degrees-of-freedom (hamiltonian-like)
quantities $h^{\prime}(x;a^{\prime})=a^{\prime}|x|^{z}$ and $h(x,y;a,b) = a|x|^z + b|y|^z$,
is there any function $f(h)$ such that {\it exact} renormalization occurs
upon reducing the degree of freedom $y$? More explicitly, is there any $f(h)$
 such that $\int_{-\infty}^{\infty} dy\; f(h) \propto f(h^{\prime})$? It is
 well known that Maxwell thought of the exponential form, i.e., $f(h)=exp(-h)$.
  Isotropy of course fixed $z=2$ (consistently with Newtonian mechanics).
  The fact that factorization occurs
   (i.e., $f(h(x,y;a,b))=f(h^{\prime}(x;a))f(h^{\prime}(y;b))$) can be
   considered, from the present viewpoint, as a simplifying aside consequence.
   Incidentally, it is worthy to note that, at Maxwell's time, his arguments
   constituted a breakthrough; indeed, the distribution  of velocities occasionally
    employed at that time was flat within some interval, and zero outside.

At this stage, let us point out that the exponential function is but the $q=1$
member of an entire family of functions, namely the $q$-exponentials
 \cite{quimicanova,borges,tsallisspringer} $f(h)=exp_q(-h) \equiv [1-(1-q)h]^{\frac{1}{1-q}}$
  {\it which, for a continuous range of values of $q$, are renormalizable, and
  which also can satisfy isotropy}. Let us more precisely define the $q$-exponential function:
\begin {equation}
e_q^t \equiv [1+(1-q)\;t]^{\frac{1}{1-q}}\;\;\;\;\;\;(t \in {\cal
R};\;q\in {\cal R})\;.
\end{equation}
For $q<1$, the $q$-exponential function vanishes for $t \le -1/(1-q)$ and continuous
and monotonically increases from $0$ to $\infty$ when $t$ increases
from $-1/(1-q)$ to $\infty$. For $q>1$, the $q$-exponential function
continuous and monotonically increases from $0$ to $\infty$ when $t$
increases from $-\infty$ to $1/(q-1)$, remaining divergent for
$t>1/(q-1)$. We easily verify that $e_1^t \equiv \lim_{q \rightarrow 1+0} e_q^t=\lim_{q
\rightarrow 1-0} e_q^t=e^t\;\;(\forall t)$.

Let us now illustrate, for a simple case, the above mentioned renormalizability:
\begin{eqnarray}
\int_{-\infty}^{\infty}&dx_2&\;  exp_q(-a_1 |x_1|^{z_1}-a_2 |x_2|^{z_2})     \nonumber\\
&=& A_2 \, exp_{q'}(-a'_1|x_1|^{z_1}) \;\;\;\;(a_1,a_2,z_1,z_2>0)
\end{eqnarray}
where
\begin{equation}
\frac{1}{1-q'}=\frac{1}{1-q}+\frac{1}{z_2},
\end{equation}
\begin{equation}
a_1^{\prime}(1-q^{\prime})=a_1(1-q)
\end{equation}
and
\begin{eqnarray}
A_2&=&\int_{-\infty}^{\infty} d\xi \; exp_q(-a_2 |\xi|^{z_2})=
\frac{2\,\Gamma\left(1+\frac{1}{z_2}\right)}{a_2^{1/z_2} } \nonumber \\
&&\times
\left\{\matrix{
\frac{\Gamma\left(1+\frac{1}{1-q}\right)}
{(1-q)^{1/z_2}
\Gamma\left(1+\frac{1}{z_2}+\frac{1}{1-q}\right)}
& \mbox{for} \,\,\,q<1
\cr
1
& \mbox{for} \,\,\,q=1
\cr
\frac{\Gamma\left(\frac{1}{q-1}-\frac{1}{z_2}\right)}
{(q-1)^{1/z_2} \Gamma\left(\frac{1}{q-1}\right)}
& \mbox{for} \,\,\,q>1 \cr  }  \right.
\end{eqnarray}
This type of renormalization remains exact for the following $N$-degrees-of-freedom
 dimensionless Hamiltonian:
\begin{equation}
h_N(\{x_i\};\{a_i\})=\sum_{i=1}^N a_i |x_i|^{z_i}
\end{equation}
We straightforwardly verify that
\begin{eqnarray}
\int_{-\infty}^{\infty}     &dx_N&\;  exp_q(-h_N(\{x_i\};\{a_i\}))  \nonumber \\
&=& A_N \, exp_{q^{\prime}}(-h_{N-1}(\{x_i\};\{a^{\prime}_i\}))\;,
\end{eqnarray}
where $q^{\prime}$ and $A_N$ are respectively given by Eqs. (3) and (5), with $z_N$
and $a_N$ replacing $z_2$ and $a_2$. Eq. (4) is generalized
 into $a_i^{\prime}(1-q^{\prime})=a_i(1-q)\;\;(i=1,\;2,...,\;N-1)$. It is
 clear that there is no major difficulty in reducing, at every step, more
 than one degree of freedom. Hamiltonian (6) with $z_i=2\;(\forall i)$ corresponds,
 of course, to $N$ free particles in one dimension, or $N/d$ particles in $d$ dimensions; isotropy
 is automatically satisfied in such cases. Also, with $z_i=2$ for $N/2$ degrees of
 freedom and $z_i=z$ for the other $N/2$ degrees of freedom, it corresponds to $N/2$
  one-dimensional anharmonic oscillators.

Let us now make the junction of the above renormalizability ideas with the nonextensive
statistical mechanics introduced a decade ago  \cite{tsallisjsp}, further implemented
 in \cite{curado,tsamepla} and applied in a considerable variety of systems, such as
  Levy \cite{levy} and correlated \cite{correlated} anomalous diffusions, peculiar
   velocities in spiral galaxies \cite{galaxy}, turbulence in electron plasma\cite{bogho},
   fully developed turbulence \cite{turbulence}, citations of scientific papers \cite{citations},
    reassociation in folded proteins \cite{bemski}, quantum entanglement \cite{rajagopal},
     and others (for recent reviews, see \cite{tsallisspringer,tsallisbjp}).
      To be more precise, this formalism addresses systems which, in one way
       or another, exhibit spatial and/or temporal long-range interactions,
        multifractal structures, dissipation, and related anomalies. The basic formal
        standpoint for thermal equilibrium (or equilibrium-like stationary states)
         consists in optimizing, under appropriate constraints, the entropic
          form $S_q=[1-\sum_ip_i^q]/[q-1]$ ($q \in {\cal R}; \; \{p_i\}$ is the
           set of probabilities associated with the microscopic states). In particular,
           for the canonical ensemble, the constraints are \cite{tsamepla} $\sum_i p_i=1$
           and $\sum_iP_i E_i=U_q$ ($\{E_i\}$ is the set of the energies of the microscopic states),
            where $P_i=p_i^q/\sum_j p_j^q$. It is clear that the $q \rightarrow 1$ limit
            reproduces the standard Boltzmann-Gibbs recipe; in particular $S_1=-\sum_ip_i \ln\;p_i$
            (we are using $k_B=1$). Also, if $A$ and $B$ are two probabilistically independent
             systems, we have that $S_q(A+B)=S_q(A)+S_q(B)+(1-q)S_q(A)S_q(B)$, hence $q=1,\;>1, \;<1$
              respectively correspond to extensive, subextensive and superextensive systems.

Optimization for arbitrary $q$ straightforwardly leads \cite{tsamepla} to
\begin{equation}
p_i=\frac{exp_q(-(E_i-U_q)/T_q)}{\overline{Z}_q},
\end{equation}
where $\overline{Z}_q=\sum_jexp_q(-(E_j-U_q)/T_q)$ with $T_q \equiv T\sum_jp_j^q$, $1/T \equiv \beta$
 being the Lagrange parameter associated with the energy constraint. As we see, the similarities
 with Boltzmann-Gibbs statistical mechanics are quite striking. For instance, it  can be verified
  that $ 1/T=\partial S_q/\partial U_q$. Also, systematically, the function $e_q^t$ and
  its inverse $\ln_q t \equiv (t^{1-q}-1)/(1-q)$ play the usual roles of the exponential
   and logarithm ones. In particular, $F_q \equiv U_q-TS_q=-T \ln_q Z_q$
   and $U_q= - \partial (\ln_q Z_q)/ \partial \beta$, where $Z_q$ is defined through
$ \ln_q Z_q= \ln_q \overline{Z}_q-\beta U_q$.

The application of this formalism to the classical Hamiltonian (6) yields
\begin{eqnarray}
\overline{Z}_q &=& \int d^N{\bf x}\;exp_q(-(h_N(\{x_i\};\{a_i\})-U_q)/T_q) \nonumber \\
 &=& I_{q,\{\frac{a_i}{\tau_q}\}}
exp_q(U_q/T_q) ,
\end{eqnarray}
where $\tau_q \equiv \left[ 1+(1-q)U_q/T_q\right] T_q $ and $ I_{q,\{b_i\}} \equiv
\int d^N{\bf x}  \; exp_q(-h_N(\{x_i\};\{b_i\})) $ is easily, though tediously, calculated from the
recursive application of Eqs. (2-5).
Analogously we obtain
\begin{eqnarray}
\tilde{ Z}_q &\equiv&
\int    d^N{\bf x} \;  \left[exp_q(-(h_N(\{x_i\};\{a_i\})-U_q)/T_q)\right]^q \nonumber \\
&=& I_{\tilde{q},\{\frac{qa_i}{\tau_q}\}}
[exp_q(U_q/T_q)]^q ,
\end{eqnarray}
where $\tilde{q} \equiv 2 - 1/q$.
By using the identity \cite{tsamepla} $\overline{Z}_q=\tilde{ Z}_q$ we can solve the present set
of implicit equations. It follows the generalized classical equipartition principle (see also \cite{yamano})
\begin{equation}
U_q=\left[\sum _{i=1}^N \frac{1}{z_i}\right]T_q\;.
\end{equation}
Consequently $\tau_q = \left[1+(1-q)\sum_{i=1}^N z_i^{-1}\right]   T_q $ . It is kind of remarkable
 that such a strongly Boltzmann-Gibbs-like equality as Eq. (11) does hold, {\it in spite of the fact
 that the involved distributions are not factorizable.}

Finally we must focus on the probability distributions: $P_q^{(N)} \propto \left(p_q^{(N)}\right)^q \propto
exp_{\tilde{q}}\left(-\frac{q\,h_N}{\tau_q}\right)$. Then, by reducing one degree of freedom (as done
in Eqs. (2-5)), we obtain
\begin{eqnarray}
P_q^{(N-1)}&=& \int dx_N \;  P_q^{(N)}   \nonumber \\
&\propto&
exp_{(2-1/q^{\prime})}\left(-q\; h_{N-1}\left(\left\{x_i\right\};
\left\{a_i/\tau_q^{\prime} \right\} \right)\right).
\end{eqnarray}
Consequently, by considering $z_i=z\;(\forall i)$ we obtain
\begin{equation}
q^{\prime} = \frac{1-q+qz}{1-q+z}
\end{equation}
and
\begin{equation}
T_q^{\prime}=\left(\frac{qz}{1-q+qz} \right) T_q \;,
\end{equation}
where we have used the fact that $T_q \propto \tau_q$ with a proportionality
coefficient which preserves, as easily verifible, the initial value of $q$,
and not the renormalized one. The results corresponding to say N anharmonic
oscillators are completely analogous.

The recurrence for $q$ has only one (double) fixed point, namely $q=1$. This
point locally is an inflexion one, but {\it globally is an attractor}
(like the tangent bifurcation of the logistic map at the entrance of
the cycle-3 window): see Fig. 1. In other words, Boltzmann-Gibbs statistical
mechanics is, as well expected, the correct one for the independent-particle
system we have focused on here. We also verify that (i) $T_q=0$ is an invariant
subspace of the renormalization group; (ii) there is no $T_q$-flow at the $q=1$
subspace, and (iii) Eq. (14) is invariant through the
transformation $T_q \rightarrow \lambda T_q$ ($\lambda >0$)
(which means that, without loss of generality, we can always
start the recurrence with say $T_q=1$). A typical two-dimensional
flow is shown in Fig. 2. Before going on, let us mention that, if
we start the recurrence with $0<q<1$, the flow in the $q,T_q$ space
is smooth and monotonic ($q$ approaches unity, and $T_q$ approaches zero).
If we start with $q$ not much larger than unity, the flow is smooth and
monotonic (both $q$ and $T_q$ increase) as long as the iterations provide
$q$-images not exceeding $1+z$ (divergence of $q^{\prime}$ in Eq. (13)).
After that, the behavior becomes physically meaningless. Indeed, the $q$-image
is quickly sent to quite negative values, then approaching $q=1$ from below.
Concomitantly, $T_q$ changes sign, and approaches zero from below. This
mathematical artifact is easy to understand; indeed, the integrals involved
in the renormalization are not defined for $q \ge 1+z$.

Since the renormalization takes $N$ into $N-1$, all fixed points necessarily correspond
to the thermodynamic limit $N \rightarrow \infty$. The same type of flow is expected to
correspond to any (classical or quantum) system with  noninteracting elements. If there
are collective {\it short-range} interactions (e.g., a classical $d$-dimensional fluid
with two-body interactions decaying as $r^{-\alpha}$, or an Ising ferromagnet with
coupling constant decaying as $r^{-\alpha}$, with $\alpha>d$), we still expect $q=1$
to be the unique (globally attractive) fixed point, but on the $q=1$ invariant subspace
there might be a finite $T_q$ unstable fixed point, $T_q=0$ and $T_q \rightarrow \infty$
being stable fixed points, respectively corresponding to the ordered and disordered phases.
{\it However}, if the interactions are {\it long-range} ones, we expect the double fixed
 point $q=1$ to split into {\it two} fixed points, namely a globally attractive one
 at $q=1$, and a globally repulsive one at $q^*>1$, whose value
should depend on $(\alpha,d)$ (naturally, $q^*$ is expected to
continuously approach unity when $\alpha/d \rightarrow 1$). See
Fig. 3. A clarification is needed: why we rather expect $q^*>1$
and not $q^*<1$? This comes from an everyday increasing evidence
in a variety of nonextensive systems (electron-positron
annihilation \cite{bediaga}, quark-gluon plasma \cite{rafelski},
granular matter \cite{granular}, cosmology \cite{ivano}, $d=1$
system of inertial classical planar rotators ferromagnetically
coupled at long distances \cite{cataniario}, among others), where
the Boltzmann exponential distribution of energies is replaced by
a long-tailed power-law, and this precisely is what occurs for
$q>1$ (see Eq. (8)). One more clarification is needed: why we
rather expect the $q=1$ fixed point,  and not the $q \ne 1$ one,
to be the (globally) attractive one? Once again, this comes from
increasing evidence \cite{ruffo} that the Boltzmann regime is the
ultimately stable one for any finite-size system (it corresponds
to the $lim_{N \rightarrow \infty} lim_{t \rightarrow \infty}$),
whereas the nonextensive regime (i.e., $q \ne1$)  emerges in the
$lim_{t \rightarrow \infty} lim_{N \rightarrow \infty}$ ordering
(see also Fig. 4 of \cite{tsallisbjp}). Moreover, the present
conjecture is consistent with the topology recently found in
\cite{fernando} for anomalous diffusion, namely a stable fixed
point at $q=1$ and an unstable one at $q=2$.

Summarizing, we have shown that the {\it renormalizability} and {\it isotropy}
principles which guided Maxwell arguments in establishing his celebrated Gaussian
 distribution of velocities is not a privilege of the exponential function, but
 it is shared by an entire power-law family of functions (the $q$-exponentials),
 which includes the standard exponential as the $q=1$ limiting case. This provides
  a remarkable bridge with nonextensive statistical mechanics, where precisely
  the $q$-exponentials play a central role (just as the exponential does within
   Boltzmann-Gibbs statistical mechanics). This observation straightforwardly
   enabled the establishment of the $q$-generalized classical principle of
    equipartition of energies. It also enabled the formulation of an exact
    renormalization group in the ($q$, temperature)- space, which shows that
    the globally stable fixed point $q=1$ (i.e., Boltzmann-Gibbs statistical
    mechanics) is a {\it double} one. This allowed the conjecture that, for
    systems including long-range interactions, this fixed point splits into
    two, one of them ($q=1$) remaining globally stable, and the other
    one (presumably $q>1$) being globally unstable. This situation
    respectively reflects the fact that the extensive or nonextensive
    thermostatistical are to be observed in the limits where the first
     to achieve infinity is the time or the size.

One of us (RSM) acknowledges S. Matteson and P. Grigolini for warm hospitality
at the Department of Physics/UNT.

\newpage
\begin{figure}
\caption{Noninteracting Hamiltonian (with $z=2$): Renormalization of $q$ (any initial
value of $q$ is ultimately attracted by $q=1$; in particular, if we start with $q>1$,
 the successive values of $q$ might increase for a few steps, but then the image is
 sent to the region $q<<1$, and then it monotonically increases up to $q=1$).
  Renormalization alternates between the curves $q^{\prime}=q$ and $q^{\prime}$
   given by Eq. (13), which exhibits a vertical asymptote at $q=z+1$.}
\end{figure}

\begin{figure}
\caption{Noninteracting Hamiltonian (with $z=2$): Renormalization group
 flow in the $(q,T_q)$ space, by starting with three different initial
 values of $q$; the initial value for $T_q$ is always taken to be unity (see the text).}
\end{figure}

\begin{figure}
\caption{Typical long-range interacting Hamiltonian: Conjectural
 projection on the $q$-axis of a two- or more-dimensional renormalization group flow.
  Two fixed points are expected, namely at $q=1$ (attractor) and at $q=q^*$ (repulsor);
   $q^*$ should approach unity when the range of the interactions decreases (for all $\alpha$
    above some critical value, $q^*$ is expected to remain equal to unity).}
\end{figure}

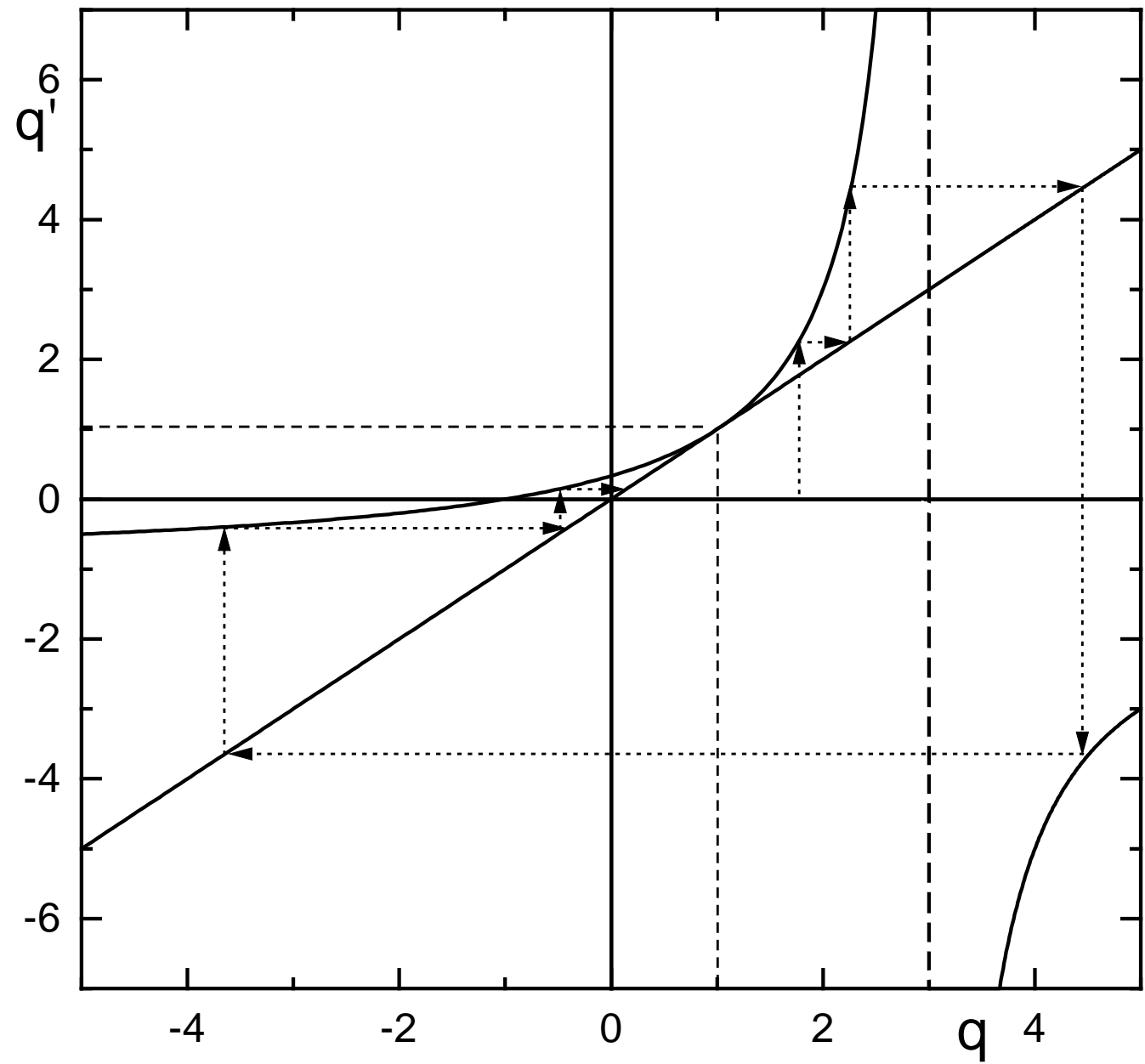
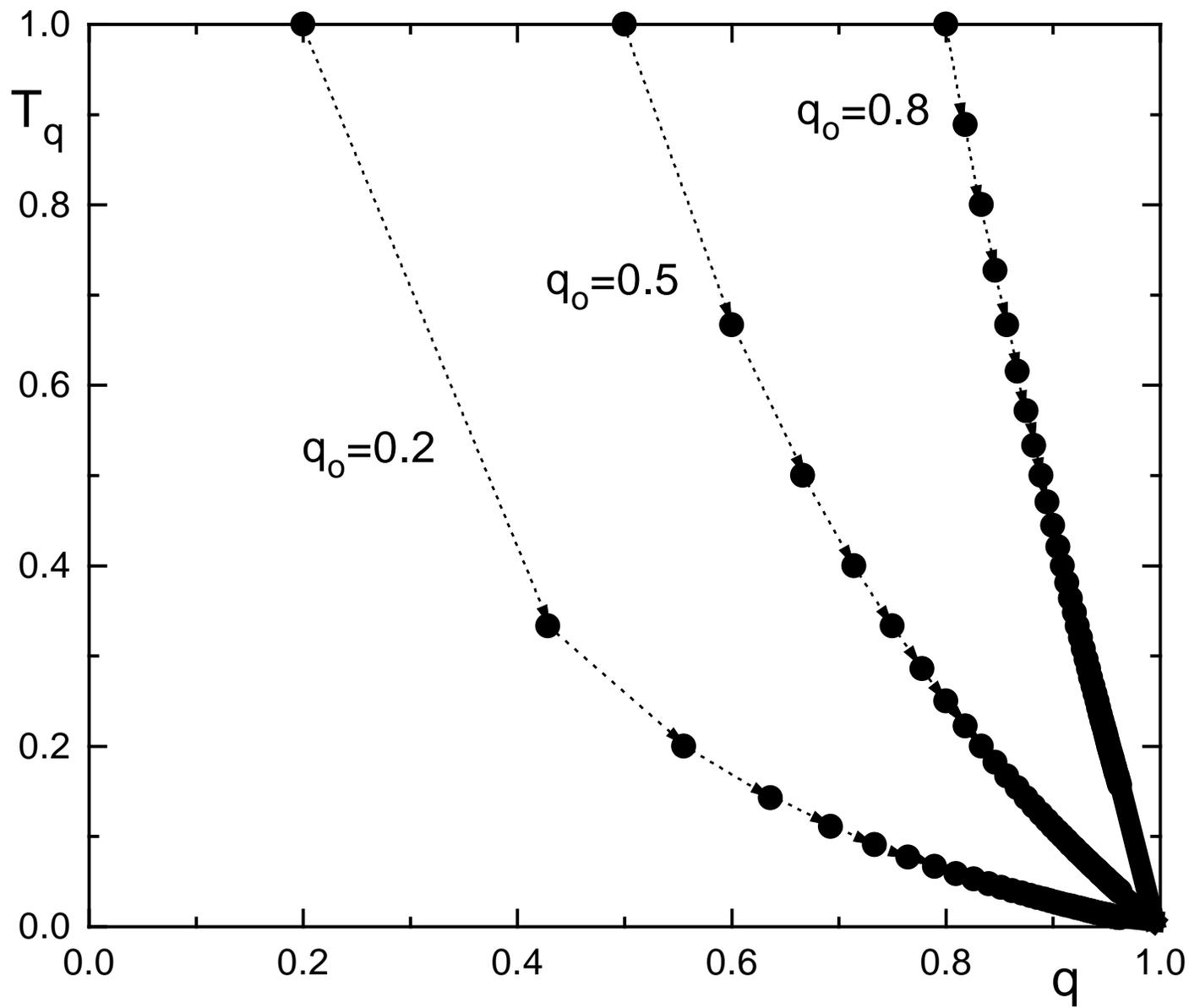
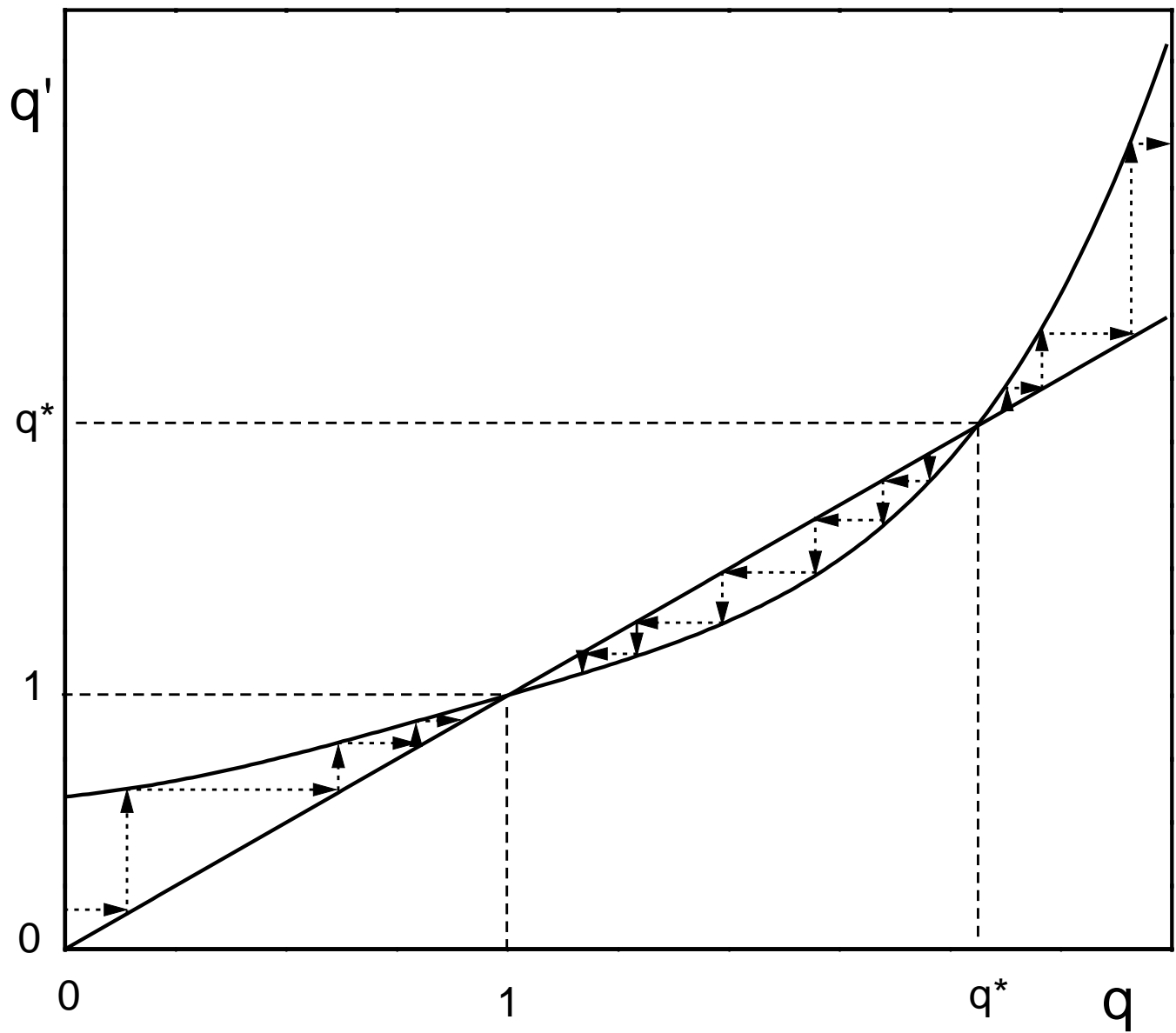
\end{document}